\documentclass[%
superscriptaddress, %
twocolumn, %
showpacs,
 amsmath,amssymb,
 aps,
prb,
floatfix, longbibliography ]{revtex4-1}
\usepackage{graphicx}
\usepackage{dcolumn}
\usepackage{bm}
\usepackage{soul,color}

\usepackage[colorlinks,linkcolor=blue,anchorcolor=blue,citecolor=blue,urlcolor=blue]{hyperref}
\begin{document}

\title{Minimum model for the electronic structure of \\
       twisted bilayer graphene and related structures}

\author{Xianqing Lin}
\affiliation{Physics and Astronomy Department,
             Michigan State University,
             East Lansing, Michigan 48824, USA}
\affiliation{College of Science,
             Zhejiang University of Technology,
             Hangzhou 310023, China}

\author{David Tom\'anek}
\email[E-mail: ]{tomanek@pa.msu.edu}
\affiliation{Physics and Astronomy Department,
             Michigan State University,
             East Lansing, Michigan 48824, USA}

\date{\today}

\begin{abstract}
We introduce a minimum tight-binding model with only three
parameters extracted from graphene and untwisted bilayer graphene.
This model reproduces quantitatively the electronic structure of
not only these two systems and bulk graphite near the Fermi level,
but also that of twisted bilayer graphene including the value of
the first magic angle, at which bands at $E_F$ flatten without
overlap and two gaps open, one above and one below $E_F$. %
{Our approach also predicts the second and third magic angle.}
The Hamiltonian is sufficiently transparent and flexible to be
adopted to other twisted layered systems.
\end{abstract}

\pacs{%
}



\maketitle


The electronic structure of graphite has been described
quantitatively as early as 1947 by Wallace~\cite{Wallace47} and
found to be dominated by $p_\perp$
orbitals~\cite{SlonczewskiWeiss58} near the Fermi level $E_F$. It
is amazing how this system continues providing surprises in the
behavior of charge carriers near $E_F$. In monolayer graphene
(MLG), described quantitatively by a one-parameter
Hamiltonian~\cite{CastroNeto2009}, backscattering of the massless
fermions
near the Dirac point $K$ in the corner of the hexagonal Brillouin
zone (BZ) is suppressed due to the Klein paradox. In bilayer
graphene (BLG) with the Bernal AB layer stacking,
the inter-layer interaction turns the linear band dispersion at
$K$ to a parabola and massless fermions in MLG to massive fermions
in BLG and graphite. Most recently, correlated
insulating~\cite{Cao2018} and unconventional
superconducting~\cite{cao2018unconventional} behavior have been
reported in magic-angle twisted bilayer graphene (TBLG).
Theoretical description of TBLG turns out to be challenging, since
unit cells in the Moir\'{e} pattern of the bilayer become
infinitely large for the general case of incommensurate
structures. An elegant solution to this problem has been provided,
treating the inter-layer interaction in a continuum model and
handling the inter-layer matrix elements in reciprocal
space~\cite{{LopesdosSantos2007},{Bistritzer12233},{LopesdosSantos2012}}.
Even though band flattening at $E_F$ and gap opening near $E_F$
have been predicted theoretically using many approaches~\cite{{Morell2010},%
{Bistritzer12233},{Trambly2012},{Moon2012},{LopesdosSantos2012},%
{Jung2014},{Cao2016},{Fang2016},{Nam2017},{Kim3364}}, none has
succeeded so far to reproduce the observed value of the (first)
magic angle $\theta_{m,1}=1.1^\circ$ accompanied by a band
flattening without band overlap
at $E_F$, opening of band gaps both below and above the flat
bands~\cite{{Cao2018},{cao2018unconventional}}, and a sharp
resistance increase at the charge neutrality point.

\begin{figure}[b!]
\includegraphics[width=1.0\columnwidth]{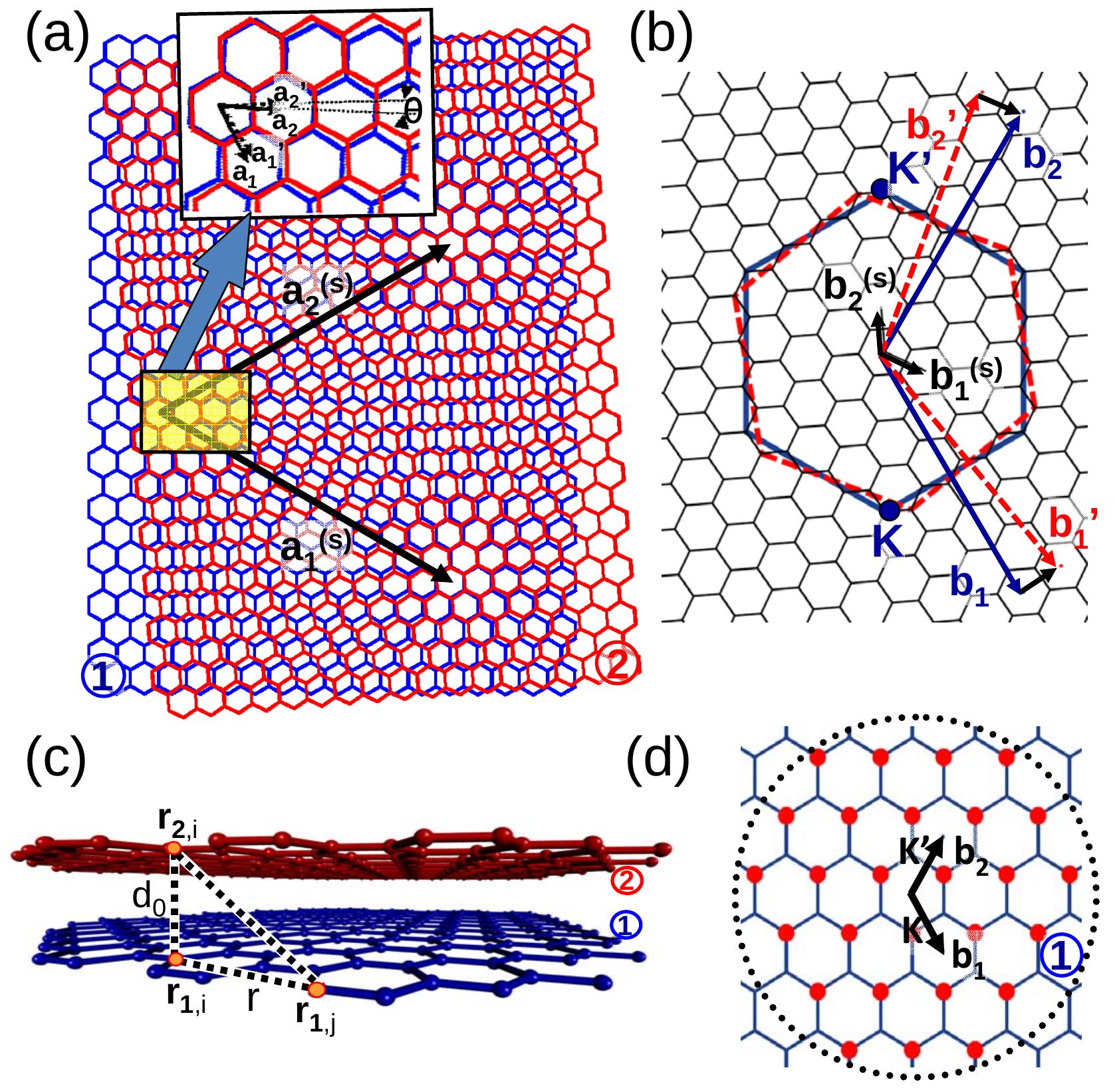}
\caption{(Color online) Schematic structure of TBLG. %
(a) Moir\'{e} superlattice formed by placing layer~2 (red),
twisted by $\theta$, on top of layer~1 (blue). The lattice vectors
${\bf{{a}_1}}$ and ${\bf{{a}_2}}$, also shown in the enlarged
inset, span the Bravais lattice of layer~1. The primed quantities
correspond to layer~2, and the superscript $(s)$ identifies the
Moir\'{e} superlattice. %
(b) Large Brillouin zone of layer~1 (blue), spanned by $\bf{b_1}$
and $\bf{b_2}$, and of the twisted layer~2 (red), spanned by
${\bf{b'_1}}$ and ${\bf{b'_2}}$. The inequivalent Dirac points $K$
and $K'$ are in the corners of the hexagonal unit cells of the
individual layers. The small hexagonal Brillouin zones tiling the
reciprocal space are spanned by ${\bf{b_1^{(s)}}}$ and
${\bf{b_2^{(s)}}}$. %
(c) Definition of interatomic distances in adjacent layers
separated by $d_0$ in perspective side view. %
(d) Brillouin zones in the reciprocal lattice of layer~1.
\label{fig1}}
\end{figure}


Here we construct a minimum tight-binding Hamiltonian with only
three parameters extracted from MLG and untwisted BLG. This
Hamiltonian reproduces quantitatively the electronic structure of
not only these two systems and bulk graphite near $E_F$, but also
that of TBLG including the values of the %
{magic angles $\theta_{m,1}$, $\theta_{m,2}$, and $\theta_{m,3}$.}
At $\theta_{m,1}$, bands at $E_F$ flatten without overlap,
two gaps open, one above and one below $E_F$. The
Hamiltonian is sufficiently transparent and flexible to be adopted
to other twisted layered systems.

As mentioned above, none of the computational approaches used so
far to describe TBLG and the role of the magic angle has succeeded
in reproducing all aspects of the observed
data~\cite{{Cao2018},{cao2018unconventional}}. An elegant
description of TBLG using the continuum model and treatment of the
inter-layer hopping in Fourier space has been introduced in
Ref.~\cite{LopesdosSantos2007},
but did not find gaps in the electronic spectrum in the range of
twist angles $\theta$ investigated.
The follow-up paper by the same authors~\cite{LopesdosSantos2012}
did find the magic angle $\theta_{m,1}{\approx}1.1^\circ$ and both
band gaps. However,
the band gap dependence on the twist angle disagrees with more
recent experimental data~\cite{Cao2018},
likely due to an inaccurate description of the inter-layer
interaction~\cite{Tang1996}. The magic angle was first predicted
in the theoretical Ref.~\cite{Bistritzer12233}, which also used
the continuum model and treated the inter-layer hopping in Fourier
space using experimentally obtained parameters. The authors
discussed the occurrence of a flat band at $\theta_{m,1}$, but did
not discuss band gaps near $E_F$. A separate calculation using the
same approach~\cite{Bistritzer12233} reproduced only one gap below
$E_F$. No band gaps were found near $E_F$ in the follow-up
study~\cite{Jung2014} based completely on \emph{ab initio} density
functional theory (DFT). The continuum model and Fourier space
treatment were abandoned in a detailed DFT study of
Ref.\cite{Fang2016} applied to commensurate structures.
The DFT results, obtained using maximally localized Wannier
functions, were mapped onto a tight-binding Hamiltonian with 18
parameters, which was diagonalized directly in the large Moir\'{e}
supercells. Even though this approach reproduces the band
flattening at the magic angle, the authors reported only one band
gap above $E_F$. %
{%
A related approach to determine the electronic spectrum, which
relies on the Hubbard model rather than DFT, has recently been
proposed~\cite{koshino18} as an extension of the
}%
initial tight-binding description of TBLG in terms of $V_{pp\pi}$
intra-layer and $V_{pp\sigma}$ inter-layer two-center hopping
integrals~\cite{Moon2012}. Whereas the initial
report~\cite{Moon2012} found band gaps only for large twist angles
beyond $\theta_{m,1}$, a follow-up study using the same
approach~\cite{Nam2017} reported crossing flat bands at the charge
neutrality point, in contrast to the observed sharp resistance
increase,\cite{Cao2018} and claimed that band gap opening at
$\theta_{m,1}$ is caused by lattice relaxation.
{%
The necessity to determine lattice relaxation to reproduce
experimental observations is computationally extremely
demanding~\cite{koshino18} and thus limits the size of the
Moir\'{e} supercells in the commensurate structure, making
prediction of higher magic angles extremely difficult. %
}%
All reported theoretical results suggest that the low-energy
electronic structure of TBLG near $\theta_{m,1}$ is rather
sensitive to the model description and the parameters.

We combined the most attractive aspects of the above theoretical
approaches
in a minimum model that is consistent with experimental
data~\cite{{Cao2018},{cao2018unconventional}}. The Hamiltonian we
propose for any graphitic system consists of an intra-layer part
$H_\parallel$ and an inter-layer part $H_\perp$. The description
we chose combines simplicity and transparency with the benefits of
previously used models while avoiding their different
shortcomings. This Hamiltonian reads
\begin{eqnarray}
H &=& H_\parallel + H_\perp \nonumber \\
&=&-\sum_{{i{\neq}j}\atop{m}} \gamma_{ij}^{mm}
(c^+_{m,i}c^{\,}_{m,j}+h.c.)
\nonumber \\
& &-\sum_{{i,j}\atop{m}}      \gamma_{ij}^{m,m+1}
(c^+_{m,i}c^{\,}_{m+1,j}+h.c.) \;.%
\label{eq1}
\end{eqnarray}
Here, $c^+_{m,i}$ is the creation and $c^{\,}_{m,i}$ is the
annihilation operator of a $p_z$ state at the atomic site $i$ in
layer $m$, with $m=1$ or $2$ for BLG. $\gamma_{ij}^{mm}$ is the
in-plane hopping integral between sites $i$ and $j$.

Typically, only nearest neighbor intra-layer hopping is considered
in $H_\parallel$. $\gamma_{<ij>}^{mm}=V^0_{pp\pi}=3.09$~eV
reproduces the Fermi velocity~\cite{CastroNeto2009}
$v_F{\approx}1{\times}10^6$~m/s in the graphene layer spanned by
lattice vectors $\bf{a_1}$ and $\bf{a_2}$, shown in
Fig.~\ref{fig1}(a), with $|{\bf{a_1}}|=|{\bf{a_2}}|=a$. The
corresponding reciprocal lattice vectors $\bf{b_1}$ and
$\bf{b_2}$, defining the BZ of the layer, are shown in
Fig.~\ref{fig1}(b).

To describe the inter-layer interaction in $H_\perp$, we first
considered an AB-stacked untwisted BLG, as illustrated in
Fig.~\ref{fig1}(c). We first consider two atoms atop each other in
adjacent layers, at the positions $\bf{r_{1,i}}$ and
$\bf{r_{2,i}}$, separated by the inter-layer distance $d_0$. The
inter-layer hopping integral between these atoms is
$t(0)=\gamma_{ii}^{1,2}=V^0_{pp\sigma}$. Next, we consider one of
the atoms moving within the layer, so that the mutual distance
vector, projected on one of the layers, becomes $|{\bf{r}}|=r>0$.
For $r$ not very large, the dominant inter-layer hopping integral
is still $V_{pp\sigma}$, scaled by the distance and corrected for
the cosine of the tilting angle~\cite{Moon2012}. It is isotropic
and can be written as
\begin{equation}
t(r) = V^0_{pp\sigma} e^{-(\sqrt{r^2+d_0^2}-d_0)/\lambda}
\frac{d_0^2}{r^2+d_0^2} \,, %
\label{eq2}
\end{equation}
where $\lambda$ modulates the cutoff of $t(r)$ at large distances.
This expression allows a flexible description of the inter-layer
interaction in regions of local AA and AB stacking as well as
in-between.

Precise observations for AB-stacked untwisted BLG provided
accurate values $a=2.46$~{\AA}, $d_0=3.35$~{\AA} and
$V^0_{pp\sigma}=0.39$~eV$=\gamma_1$ in standard graphite
notation. %
{Using ${\lambda}=0.27$~{\AA}, we could furthermore reproduce the
well-established band structure of AA- and AB-stacked BLG. This
value of ${\lambda}$ also yielded }
$\gamma_3=\gamma_4=0.11$~eV for neighbors in adjacent layers with
$r=a/\sqrt{3}$ in very good agreement with experimental
data~\cite{{BLGparam13},{BLGparam14},{BLGparam16}}. All parameters
needed to reproduce the electronic structure of MLG, BLG, graphite
and TBLG are listed in Table~\ref{table1}. As we will show,
Hamiltonian (\ref{eq1}) also reproduces the magic angle
$\theta_{m,1}{\approx}1.1^\circ$, band flattening without band
overlap at $E_F$, opening of two gaps, one below and one above
$E_F$, and band gap reduction for twist angles deviating from
$\theta_{m,1}$.

\begin{table}[t]
\caption{%
Band-structure parameters of graphitic systems.%
}%
{\begin{tabular}{lcccccccccc} \hline\hline
Quantity & $a$          &            & $d_0$      &            %
   &$V^0_{pp\pi}$&            &$V^0_{pp\sigma}$&            & $\lambda$ \\
\hline
Value    & $2.46$~{\AA} &\hspace{5mm}& 3.35~{\AA} &\hspace{5mm}%
   & 3.09~eV     &\hspace{5mm}& 0.39~eV        &\hspace{5mm}& $0.27$~{\AA} \\
\hline\hline
\end{tabular}%
}%
\label{table1}
\end{table}

In the following, we will describe a TBLG initially formed as an
AA stacked BLG, where the top layer~2 has been twisted
counterclockwise by the angle $\theta$ with respect to the bottom
layer~1, as seen in top view in Fig.~\ref{fig1}(a). The honeycomb
lattice of a graphene layer consists of a triangular Bravais
lattice with a two-atom basis. The vectors spanning the Bravais
lattice of the bottom layer~1 are
${\bf{{a}_1}}=a(\sqrt{3}/2,-1/2)$ and ${\bf{{a}_2}}=
a(\sqrt{3}/2,1/2)$ in Cartesian coordinates. The positions of the
two basis atoms A and B in the unit cell, which form the
sublattices A and B, are
${\bf{{\tau}_A}}=({\bf{{a}_1}}+{\bf{{a}_2}})/3$ and
${\bf{{\tau}_B}}=2({\bf{{a}_1}}+{\bf{{a}_2}})/3$. The Bravais
lattice vectors spanning the twisted upper layer~2 are
${\bf{{a'}_1}}$ and ${\bf{{a'}_2}}$ and the basis vectors spanning
the sublattices are ${\bf{{\tau}_{\alpha}'}}$. The reciprocal
lattice of the bottom layer~1, spanned by ${\bf{b}_1}$ and
${\bf{b}_2}$, is shown in Fig.~\ref{fig1}(d).

For commensurate TBLG lattices, we can use the index $(M,N)$ to
define the twist angle $\theta$ and the Moir\'{e}
supercell~\cite{Fang2016}. Incommensurate lattices can still be
approximated by a commensurate lattice with a specific index
$(M',N')$ and ${\theta}'{\approx}\theta$, albeit with possibly
very large supercells. The reciprocal lattice of the $(N+1,N)$
TBLG with a small twist angle, shown in Fig.~\ref{fig1}(b), is
spanned by the vectors ${\bf{b_1^{(s)}}} =
{\bf{b}_2}-{\bf{b'}_2}$, and ${\bf{b_2^{(s)}}} =
({\bf{b'}_1}+{\bf{b'}_2})-({\bf{b}_1}+{\bf{b}_2})$, where
${\bf{b}_i}$ and ${\bf{b'}_i}$ with $i=1,2$ are reciprocal lattice
vectors of the bottom and the top layer, respectively.


In the following, we will focus on a TBLG lattice with small twist
angles near the observed magic angle
$\theta_{m,1}{\approx}1.1^\circ$. Whether commensurate or
incommensurate, such a lattice can be described or approximated by
a commensurate lattice with a large Moir\'{e} supercell and the
electronic structure can be obtained to a good accuracy using the
continuum method. In this approach, the low-energy wavefunctions
can be expanded in the Bloch basis of the bottom layer~1 and the
twisted top layer~2 near the Dirac point, which are defined as
\begin{eqnarray}
|\psi_{1,\alpha}({\bf{k}})\rangle &=& \frac{1}{\sqrt{N}} %
   \sum_{{\bf{R}}} e^{i{\bf{k}}{\cdot}({\bf{R}}+{\bf{\tau}_{\alpha}})} %
   |{\bf{R}}+{\bf{\tau}_{\alpha}}\rangle , \nonumber \\
|\psi_{2,\alpha}({\bf{k}})\rangle &=& \frac{1}{\sqrt{N}} %
   \sum_{{\bf{R}}'} e^{i{\bf{k}}{\cdot}({\bf{R'}}+{\bf{\tau}'_{\alpha}})} %
   |{\bf{R'}}+{\bf{\tau}'_{\alpha}}\rangle \,.
\label{eq3}
\end{eqnarray}
Here, the index $\alpha$ denotes the $A$ or $B$ sublattice, and
the Wannier function $|{\bf{R}}+{\bf{\tau}_{\alpha}}\rangle$ is
the $p_z$ orbital at that site. To discuss the value range of
$\bf{k}$, we refer to Fig.~\ref{fig1}(b) depicting the large
hexagonal Brillouin zone of layer~1 spanned by $\bf{b_1}$ and
$\bf{b_2}$ and the counterparts for the twisted layer~2, and the
smaller Brillouin zones of the Moir\'{e} superlattice, spanned by
${\bf{b_1^{(s)}}}$ and ${\bf{b_2^{(s)}}}$. In the vicinity of the
Dirac point $K$ of layer~1 and its counterpart in twisted layer~2,
we can express
${\bf{k}}={\bf{k}}^{(s)}+{\bf{k}}_0+{\bf{G}^{(s)}}$, where
${\bf{k}}^{(s)}$ is a $\bf{k}$-point in the supercell BZ in the
center of the BZ of the monolayers and ${\bf{k}}_0$ is the center
of one of the supercell BZs containing $K$ of layer~1 in their
corners. ${\bf{G}^{(s)}}$ is a reciprocal lattice vector of the
superlattice, given by
${\bf{G}^{(s)}}=n_1~{\bf{b_1^{(s)}}}+n_2~{\bf{b_2^{(s)}}}$ with
small integers $n_1$ and $n_2$ typically in the range
$-4{\leq}n_i{\leq}+4$.

Defining ${\bf{k_1}}={\bf{k}}^{(s)}+{\bf{k}}_0+{\bf{G_1}^{(s)}}$
and ${\bf{k_2}}={\bf{k}}^{(s)}+{\bf{k}}_0+{\bf{G_2}^{(s)}}$, the
intra-layer Hamilton matrix elements are given in the Bloch basis
by
\begin{equation}
  \langle \psi_{m,\alpha}({\bf{k_1}})|H| \psi_{m,\beta}({\bf{k_2}}) \rangle =
  H_{m,\alpha \beta}({\bf{k_1}}) \delta_{{\bf{G_1}^{(s)}},{\bf{G_2}^{(s)}}}\,,
\label{eq4}
\end{equation}
with $m=1,2$ defining the layer and $\alpha$ the sublattice. The
on-site energy for both layers is set to be zero, so the diagonal
matrix elements of the Hamiltonian are
$H_{m,\alpha\alpha}({\bf{k}})=0$. For the two layers 1 and 2, the
off-diagonal matrix elements of the Hamiltonian are given by
\begin{eqnarray}
H_{1,AB}({\bf{k}}) = -V^0_{pp\pi} %
   \sum_{j=1}^{3} e^{i{\bf{k}}{\cdot}{\bf{\rho_j}}} \,, \nonumber\\
H_{2,AB}({\bf{k}}) = -V^0_{pp\pi} %
   \sum_{j=1}^{3} e^{i{\bf{k}}{\cdot}{\bf{{\rho}'_j}}} \,,
\label{eq5}
\end{eqnarray}
where $V^0_{pp\pi}$ is the intra-layer nearest-neighbor hopping
term. ${\bf{\rho_j}}$ are the vectors connecting sublattice A
sites to their three nearest neighbors in sublattice B in layer~1.
${\bf{{\rho}'_j}}$ are the corresponding nearest-neighbor vectors
in the twisted layer~2. The Hamiltonian is Hermitian, so
$H_{m,BA}({\bf{k}}) = H_{m,AB}^*({\bf{k}})$ for $m=1,2$.

\begin{figure}[t!]
\includegraphics[width=1.0\columnwidth]{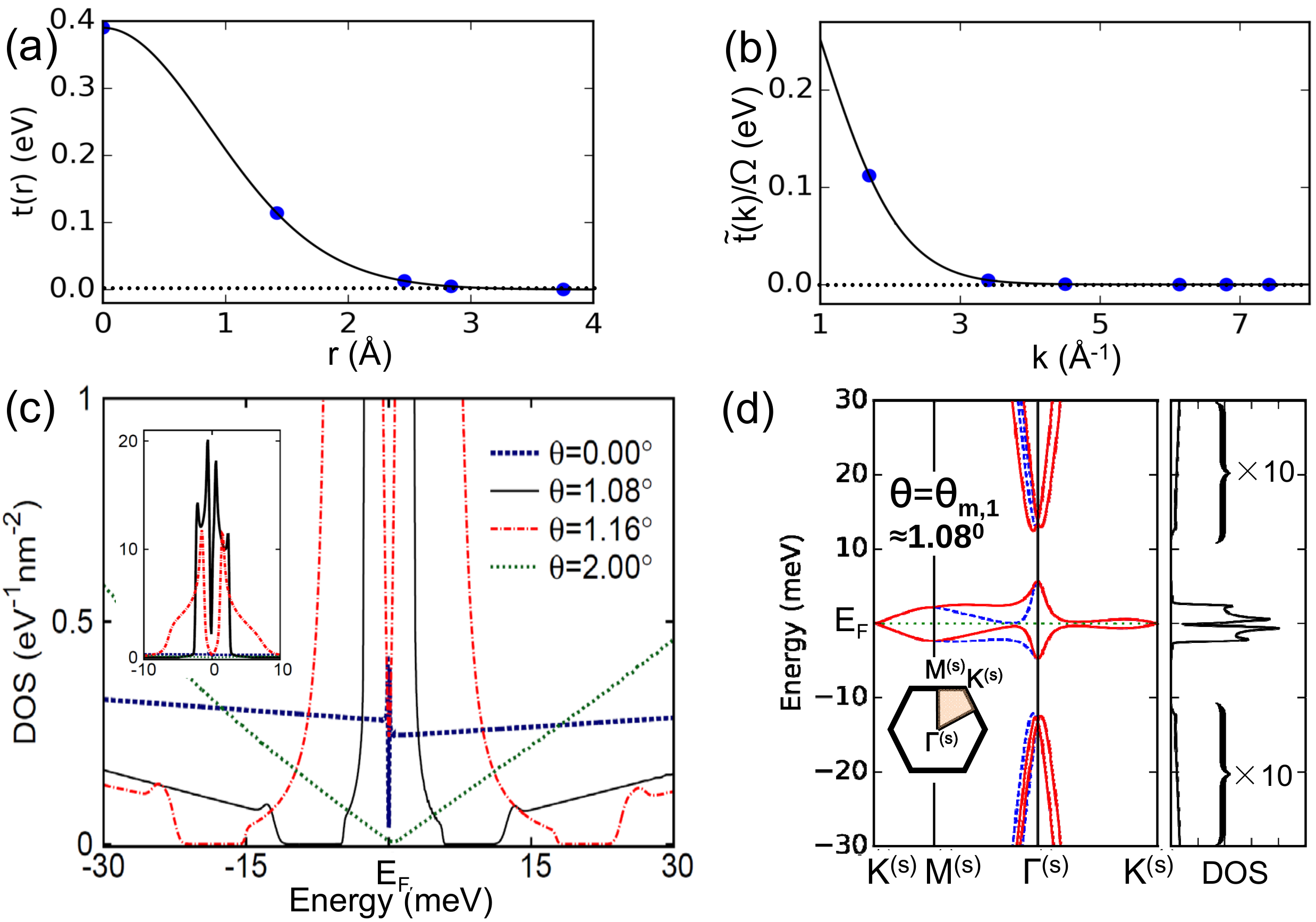}
\caption{(Color online) (a) The inter-layer hopping integral
$t(r)$, defined in Eq.~(\ref{eq2}), and %
(b) its Fourier transform $\tilde{t}(k)$. The filled circles in
(a) represent possible $r$ values in AB-stacked BLG. %
(c) Electronic density of states (DOS) of TBLG near $E_F$ for
different twist angles $\theta$. The DOS for $\theta=0$ represents
an untwisted BLG. %
(d) The electronic band structure of TBLG along high-symmetry
lines of the Moir\'{e} superlattice (left) and the corresponding
DOS (right) for $\theta$ near the
%
{(first) }%
magic angle $\theta_{m,1}{\approx}1.08^\circ$. The red (dashed
blue) lines represent bands with the valley index $K$ ($K'$)
defined in Fig.~\ref{fig1}(d). The DOS below and above $E_F$ is
multiplied by $10$. The BZ of the superlattice is shown as an
inset in the band structure. \label{fig2}}
\end{figure}

To describe the inter-layer coupling in an effective, approximate
way, we fist consider the atomic distribution in a 2D graphene
layer to be continuous uniform. In that case, the 2D Fourier
transform of $t({\bf{r}})$ is given by
\begin{equation}
  \tilde{t}({\bf{k}}) = \int e^{-i{\bf{k}}{\cdot}{\bf{r}}} t({\bf{r}}) d^2r \,.
\label{eq6}
\end{equation}
Since $t({\bf{r}})$ is isotropic, Eq.~(\ref{eq6}) can be
transformed to a 1D integral
%
\begin{equation}
  \tilde{t}(k) = 2\pi \int_0^{\infty} r t(r) J_0(kr) dr \,,
\label{eq7}
\end{equation}
where $J_0$ is a Bessel function and the Fourier transform is also
isotropic in the reciprocal space. The radial dependence of the
inter-layer hopping integral $t(r)$ is shown in Fig.~\ref{fig2}(a)
and it Fourier transform $\tilde{t}(k)$ is shown in
Fig.~\ref{fig2}(b).

 \begin{figure}[t!]
 \includegraphics[width=1.0\columnwidth]{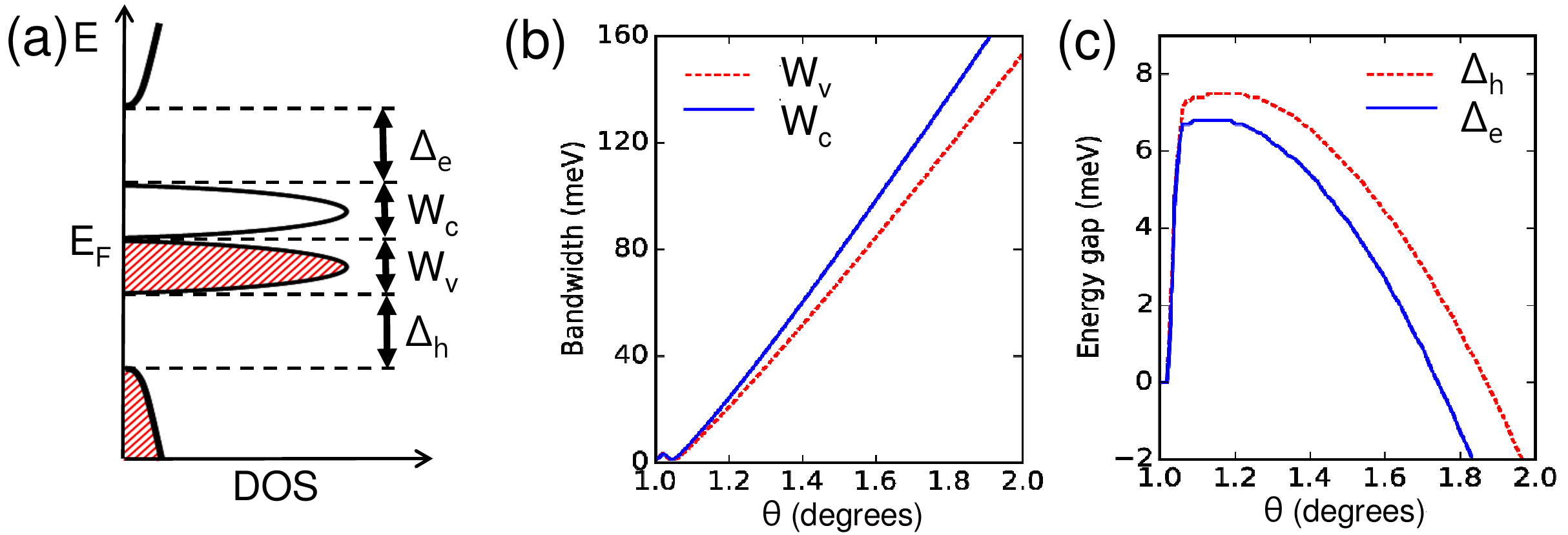}
 \caption{(Color online) Electronic structure of TBLG near the
 magic angle $\theta_{m,1}{\approx}1.1^\circ$. (a) Schematic
 electronic structure near the charge neutrality point. The flat
 band splits into two narrow valence bands of width $W_v$ and two
 narrow conduction bands of width $W_c$. A band gap of width
 $\Delta_h$ opens on the hole side below $E_F$ and a gap of
 width $\Delta_e$ opens on the electron side above $E_F$. %
 (b) $W_v$ and $W_c$ as a function of the twist angle $\theta$. %
 (c) $\Delta_h$ and $\Delta_e$ as a function of $\theta$.
 \label{fig3}}
 \end{figure}

For TBLG with a small twist angle, where the continuum model is
justified, the inter-layer Hamilton matrix elements can be
evaluated and expanded in the reciprocal space
as~\cite{Bistritzer12233}
\begin{eqnarray}
  \langle \psi_{1,\alpha}({\bf{k_1}})|H|\psi_{2,\beta}({\bf{k_2}}) \rangle = \nonumber \\
  \sum_{{\bf{G}}} \frac{\tilde{t}({\bf{k_1}}+{\bf{G}})}{\Omega} %
  e^{i({\bf{G}}{\cdot}{\bf{\tau}_{\alpha}}-{\bf{G}}{\cdot}{\bf{\tau}_{\beta}})}
  \delta_{{\bf{k_2}}-{\bf{k_1}},{\bf{G}}-{\bf{G'}}} \,.
\label{eq8}
\end{eqnarray}
Here, ${\bf{G}}$ are reciprocal lattice vectors of the untwisted
graphene layer~1, ${\bf{G'}}$ are the corresponding vectors of the
twisted layer~2, and $\Omega$ is the area of the graphene unit cell. %
${\bf{k_1}}$ and ${\bf{k_2}}$  have been defined earlier for use
in the intra-layer Hamilton matrix elements in Eq.~(\ref{eq4}).

In the expansion over the reciprocal lattice of layer~1, we found
that 27 ${\bf{G}}$-vectors, indicated by orange circles in
Fig.~\ref{fig1}(d), are necessary to reach convergence of the
electronic structure due to the larger extent of the
Fourier-transformed inter-layer hopping integral $\tilde{t}(k)$
associated with our small value of $\lambda$. In previous
studies~\cite{LopesdosSantos2007,Bistritzer12233}, only 3 small
${\bf{G}}$-vectors have been used for the expansion in
Eq.~(\ref{eq8}). Even in this restricted expansion, the authors
probed the relevant part of reciprocal space near the Dirac point
${\bf{K}}$, since $|{\bf{K}}+{\bf{G}}|$ is close to $|{\bf{K}}|$.
In the expansion of TBLG wavefunctions, we use a $9{\times}9$ grid
of ${\bf{G}^{(s)}}$-vectors for each value of ${\bf{k}}^{(s)}$.

Recent observations~\cite{{Cao2018},{cao2018unconventional}}
suggest that the (first) magic angle in TBLG, accompanied by a
band flattening and a sharp resistance increase at the charge
neutrality point, caused by vanishing band overlap, occurs at
$\theta_{m,1}{\approx}1.08^\circ$. Even though the magic angle
structure is likely incommensurate, nearby twist angle values may
be obtained considering commensurate TBLGs with index $(N+1,N)$.
Since the BZ collapses to zero in incommensurate structures, only
the DOS and not the band structure can be provided. The DOS of
TBLG with $\theta$ in the range from $0^\circ{-}2^\circ$, with
emphasis on the first magic angle $\theta_{m,1}$, is shown in
Fig.~\ref{fig2}(c) and as movie in the Supporting
Material~\cite{TBLG18-SM}. %
{Also presented in the Supporting Material~\cite{TBLG18-SM} is the
calculated DOS near the second magic angle
$\theta_{m,2}{\approx}0.47^\circ$ and the third magic angle
$\theta_{m,3}{\approx}0.28^\circ$. These values agree well with
previously reported values~\cite{Bistritzer12233}
$\theta_{m,2}{\approx}0.50^\circ$ and
$\theta_{m,3}{\approx}0.35^\circ$. }
The incommensurate structure with the magic angle $\theta_{m,1}$
can be approximated by a TBLG with index $(31,30)$ and twist angle
$\theta=1.08455^\circ$.
For this commensurate structure, we present both the band
structure $E(k)$ and the DOS in Fig.~\ref{fig2}(d). We notice that
at $\theta_{m,1}$, the flat band splits into valence and
conduction sub-bands originating from $K$ and $K'$ valleys shown
in Fig.~\ref{fig1}(b). These bands do not overlap at
$\theta_{m,1}$, providing an explanation for the sharp resistance
increase at the charge neutrality point.

The TBLG DOS near $\theta_{m,1}$ is shown schematically in
Fig.~\ref{fig3}(a). Below $E_F$, two flat valence bands of width
$W_v$ are separated by a hole gap of width $\Delta_h$ from
lower-lying occupied states. Above $E_F$, two flat conduction
bands of width $W_c$ are separated by an electron gap of width
$\Delta_e$ from higher occupied states. As seen in
Figs.~\ref{fig3}(b) and \ref{fig3}(c), the minimum values $W_v$
and $W_c$ with the bands not overlapping and no gaps above or
below $E_F$ occur near $\theta_{m,1}$. According to
Fig.~\ref{fig3}(c), even a small increase of $\theta$ beyond
$\theta_{m,1}$ opens gaps above and below the flat band. Even
though $\Delta_h>\Delta_e$ in general, both gaps decrease in size
with increasing value of $\theta$ and eventually close for
$\theta{\agt}1.7^\circ$. As seen in Fig.~\ref{fig2}(c), the DOS of
TBLG with $\theta=2^\circ$ shows no indication of any band gap or
a flat band.


In our minimum description, all parameters listed in
Table~\ref{table1} have well-established values based on
experimental observation. The only variable that required a
judicious choice was that of the decay length $\lambda$. At
$\theta_{m,1}$, the minimum values of $W_v$ and $W_c$ and thus the
minimum width of the flat band $W_{fb}{\approx}1.9$~meV was
obtained using $\lambda=0.21$~{\AA}. In this case, overlap of the
narrow valence and conduction bands along the $G^{(s)}-M^{(s)}$
direction yielded a rather large DOS at $E_F$, which is
inconsistent with the observed high resistance at the neutrality
point. We found $W_{fb}$ to increase for both $\lambda<0.21$~{\AA}
and $\lambda>0.21$~{\AA}. The narrowest flat band with
$W_{fb}{\approx}4.7$~meV and no overlap between the flat valence
and conduction bands occurred for $\lambda=0.27$~{\AA}. This value
has been used throughout our study.

In conclusion, we introduced a minimum tight-binding Hamiltonian
with only three parameters extracted from graphene and untwisted
bilayer graphene. We found that this Hamiltonian reproduces
quantitatively the electronic structure of not only these two
systems and bulk graphite near the Fermi level, but also that of
twisted bilayer graphene including the value of the first magic
angle, at which bands at $E_F$ flatten without overlap and two
gaps open, one above and one below $E_F$. %
{Our approach also predicts the second and third magic angle.}
The Hamiltonian is sufficiently transparent and flexible to be
adopted to other twisted layered systems.


\label{Acknowledgments}
\begin{acknowledgments}
D.T. acknowledges financial support by the NSF/AFOSR EFRI 2-DARE
grant number EFMA-1433459. X.L. acknowledges support by the China
Scholarship Council. We thank Dan Liu for useful discussions.
Computational resources have been provided by the Michigan State
University High Performance Computing Center. %
\end{acknowledgments}


%

\end{document}